\documentstyle[sprocl,epsf]{article}
\scrollmode
\begin{document}

\newcommand{\al}{\mbox{$\alpha$}}
 \newcommand{\be}{\mbox{$\beta$}}
 \newcommand{\th}{\mbox{$\theta$}}
\newcommand {\mod}[1] {\mbox{$ \vert #1 \vert $}}
 \newcommand{\ga}{\mbox{$\gamma$}}
 \newcommand{\ep}{\mbox{$\epsilon$}}
\newcommand{\Ga}{\mbox{$\Gamma$}}
 \newcommand{\Up}{\mbox{$\Upsilon$}}
\newcommand{\de}[1] {\mbox{$ \delta (#1) $ }  }
 \newcommand{\om}{\mbox{$\omega$}}
\newcommand{\bec}[1] {\begin{equation}\label{#1} }
\newcommand{\eec} {\end{equation} }
 \newcommand{\De}{\mbox{$\Delta$}}
\newcommand{\ti}[1]{\mbox{$ \tilde{#1} $} }
\newcommand{\La}{\mbox{$\Lambda$}}

\title{ Constraints on heavy meson form factors derived from QCD
analyticity, unitarity and heavy quark spin symmetry }
\author{Cosmin Macesanu}
\address{Department of Physics and Astronomy,University of Rochester
Rochester, NY 14627 USA}
\maketitle \abstracts {
Using the analytic properties of two-point functions in QCD, as well as
unitarity, bounds on the $B$ meson form factor $F(q^2)$ can be derived.
Heavy quark spin symmetry, correctly taken into account, is shown to
 improve  these bounds significantly.   }

\section{Introduction}
An evaluation of 
the Cabibo-Kobayashi-Maskawa matrix element $V_{cb}$ can be obtained by
studying the weak decays of $B$ mesons to $D$ mesons: $\bar{B} \to
D l\bar\nu$. The amplitude for this process is:
\bec{bd}      
 M_{fi} = \left[ \bar{\psi}_l \frac{-ig}{2\sqrt{2}}\  \ga ^\mu
(1-\ga_5)
\psi _{\bar{\nu} _l} \right] \frac{i(-g_{\mu \nu} + q_\mu q_\nu /M_W^2)}
{q^2-M_W^2}\ \times 
\eec
$$ \times V_{cb} <D(p') \vert \bar{c} \frac{-ig}{2\sqrt{2}}\
\ga ^\mu (1-\ga _5) b \vert B(p)>$$
The corresponding total cross-section can be taken from experiment. The
leptonic vertex can be computed perturbatively;  to find $V_{cb}$ all
there is yet to
do is to find a way to evaluate the hadronic matrix element.
 
We can do so by taking advantage of the fact that $B$ and $D$ are
heavy mesons. It was shown \cite{isgur}  that the form-factors which
describe such
transitions are, in the infinite mass limit, expressed through one
universal function: the Isgur-Wise function. Using the heavy quark
effective theory \cite{review} (HQET), finite mass
 corrections suited to each particular case can be computed. 

 In this paper, we will concern ourselves with the following matrix
element:
\bec{fff}
 < B(p')\vert \bar{b} \ga ^\mu b \vert B(p)>= (p+p')^\mu F(q^2)
\mbox{~~~}
 q=p-p'
\eec

Specifically, we will try to obtain bounds on the slope (charge radius)
 of the relevant
form factor $F$ near zero momentum transfer :  
 \bec{chrad} \rho^2=
\frac{t_0}{2}\ \left[ \frac{dF}{d q^2} \right]_{q^2=0} 
, t_0=  4m_B^2
\eec 
 Thus,
information on the Isgur-Wise function can be obtained; actually, to a
good approximation $F$ coincides with this function. 

 The method applied in the following makes use of the analyticity and
unitarity of the theory.  A dispersion relation for the vacuum
polarization function it is written. A Lehmann spectral reprezentation
can be used for the imaginary part of this
function. Keeping in the spectral sum 
only the contribution of certain states we get an integral inequality
which relates the form-factors along the unitarity cut (physical region
for pair production)
and the value of the function far from this region.  The spin symmetry
results of HQET are used in order to get a better inequality. Finally,
applying standard techniques related to vector-valued analytic functions
\cite{duren}
the required constraints on (\ref{chrad}) are derived. 

  In the next section we outline the method, using as an example the
simplest case (taking into account only the contribution of $B-\bar{B}$
states in the spectral sum). In the following one, we address the problem
of including the $B-\bar{B}^*$ and $B^* -\bar{B}^*$ states; for this, the
results of HQET have to be correctly taken into account. Finally, we
present some numerical results and a short discussion. 

\section{Description of the method}
Let's consider the vacuum polarization tensor:
$$ \Pi^{\mu\nu}(x)=i<0 \vert T \{ V^\mu (x) V^\nu (0) \} \vert 0> 
, V^\mu = \bar{b} \ga ^\mu b $$
 Going into momentum space, we get the invariant amplitude $\Pi(q^2)$ :
 $$\Pi^{\mu\nu}(q)=(q^\mu q^\nu- q^2 g^{\mu\nu} ) \Pi(q^2) $$
 The analytic properties of this function (in the complex $q^2$ plane) are
easy to derive. In the spectral sum
\bec{impi1}
  Im \Pi(q^2)= \frac{-1}{3q^2}\ \pi (2\pi)^3 \times
\eec
$$ \sum_{\Ga} \delta({\bf q-p}_{\Ga} ) \de{
\mod{q^0} -E_{\Ga} } <0 \mod{V^\mu (0)} \Ga><\Ga \mod{V_\mu (0)} 0>  $$
only states which contain a $b$ and a $\bar{b}$ quark will contribute.

Such states can be uniparticle states: $\Up_1, \Up_2, \Up_3, \Up_4$
 (bound states of the $b-\bar{b}$ pair) or multiparticle states : a
$B-\bar{B}$,$B-\bar{B}^*$ or $B^*-\bar{B}^*$ pair (a $B$ is a bound state
of the heavy quark $\bar{b}$ and one of the light quarks $u, d$ or $s$;
the difference between $B$ and $B^*$ is in their internal quantum numbers;
thus $B$ is a pseudoscalar particle, while $B^*$ is a vectorial one).

From the above relation, as well as the masses of these particles, it can
be seen that the function $\Pi$ will have three poles in $q^2=m_{\Up_1} ^2,
m_{\Up_2} ^2$ and $m_{\Up_3} ^2$ and a cut starting from $q^2=t_0=4m_B ^2$
( $m_{\Up_4} ^2 > t_0$, so this pole is covered by the cut).
Therefore, we can write a dispersion relation for the derivative of $\Pi$
:
\bec{disprel}
 \frac{d\Pi(q^2)}{dq^2}\ = \Pi'(q^2)= \frac{1}{\pi}
\int_{m_{\Up_1} ^2 -\epsilon }^{\infty}
  \frac{Im\Pi(t)}{(t-q^2)^2}dt
  \eec
for $q^2$ on the real axis and $q^2 < m_{\Up_1} ^2 $.

 Furthermore, it can be shown that the contribution of a certain type of
particles in the spectral sum is positive; so , keeping only the contribution of $B-\bar{B}$ states in  (\ref{impi1})
 we can write an inequality like:
\bec{impi2}
 Im \Pi(q^2) \geq \frac{-1}{3q^2}\ \pi (2\pi)^3 \times
 \eec
$$  \sum_{\Ga(B\bar{B}  )} \delta({\bf q-p}_{\Ga} ) \de{\mod{q^0} 
-E_{\Ga} }
<0 \mod{V^\mu } B\bar{B} ><B\bar{B}\mod{V_\mu  } 0> $$
 The matrix elements appearing in this inequality can be expressed with
the form factor defined in (\ref{fff}); thus we get :
\bec{im1}  Im \Pi(q^2=t) \geq
{n_f \over 48\pi} \left(1-{t_0\over t}\right)^{3/2} \mod{F(t)}^2 \th
(t-t_0) 
\eec
where $n_f=3$ comes from the fact that we have three different types of
$B$ mesons (we presume that the form factor is the same for $B_u$ , $B_d$
and $B_s$)
; inserting in (\ref{disprel}) :
 \bec{inec1}
  \Pi'(q^2) \geq
 {1\over 16\pi^2} \int_{t_0}^{\infty}
  \frac{(t-t_0)^{3/2}}{t^{3/2}(t-q^2)^2}\ \mod{F(t)}^2 dt
  \eec

 In the above relation, there is only one unknown quantity : the $F$ form
factor ($\Pi'(q^2)$ can be computed perturbatively for $q^2 \ll m_{\Up_1}^2$).
  To get information from (\ref{inec1}) on the charge radius
(\ref{chrad}), we apply the following procedure:

Defining 
\bec{idef}
I={1\over 16\pi^2 \Pi'(q^2) } \int_{t_0}^{\infty}
  \frac{(t-t_0)^{3/2}}{t^{3/2}(t-q^2)^2}\ \mod{F(t)}^2 dt
  \eec
we perform a conformal mapping:
\bec{mapp}
 z= \frac{ \sqrt{t_0-t}-\sqrt{t_0} } { \sqrt{t_0-t}+\sqrt{t_0} } \
 \eec
which brings the $t=q^2$ plane inside the unit circle; the threshold
$t_0$ goes into $z=-1$, the points on the physical cut go onto the
circumference of the circle ; (\ref{idef}) becomes 
\bec{i2}
I = \int_{0}^{2\pi} \frac{d\th}{2\pi}\ \mod{\Phi(z)F(z)}^2
\eec
  with $z=e^{i\theta}$ and
\bec{phidef}
 \Phi(z)= \frac{1}{16}\
  \sqrt{\frac{1}{2\pi t_0 \Pi'(0) } } \ (1+z)^2 \sqrt{1-z}
  \eec
in the particular case when $q^2=0$.

 Second, we get rid of the singularities of the integrand in the unit
disk. It can be shown \cite{bart} that the form factor $F$ also has poles
at the square masses of the three $\Up$ particles;  in the $z$ plane, they
will appear as simple poles $z_1, z_2, z_3$ somewhere on the real axis
between $z=-1$ and $z=0$. To make them disappear,  multiply the
integrand in (\ref{i2}) by the so-called Blaschke functions:  
\bec{blasc}
 B(z)= \prod_{i=1}^{3} \frac{z-z_i}{1-zz_i^*}
 \eec
  which have zeroes in $z_1, z_2, z_3$, and, moreover, $\mod{B(z)}=1$
on the unit circle $\mod{z}=1$.
\footnote{ if we'd knew the residues of $F$ in its poles we would be able 
to use a better method - in essence, substract the singularities, instead of
multiplying them out; see \cite{raf1} } 
 Then :
 \bec{i3}
I =  \int_{0}^{2\pi} \frac{d\th}{2\pi}\ \mod{\Phi(z)F(z)B(z)}^2
\eec
 the integrand being an analytic function.

Finally, let's consider the following expression :
$$ I=\frac{1}{2\pi} \int_{0}^{2\pi}   \mod{ G(z)}^2 d\th \ ,\ z=e^{i\th} \
 . $$
If $G$ is analytic inside the unit disk we can expand it in a power
series:
$$ G(z=e^{i\th}) = c_0 + c_1e^{i\th} + c_2 e^{2i\th} + \ldots $$
and actually perform the integration :
$$ I= \mod{c_0}^2 +  \mod{c_1}^2+  \mod{c_2}^2 + \ldots $$
 so that  $1 \geq I$ implies :
$$ 1 \geq \mod{c_0}^2 +  \mod{c_1}^2 =\mod{G(0)}^2 +  \mod{G'(0)}^2 $$

This is a Schur-Caratheodory type inequality; in our case it reads:
\bec{qin}
 1 \geq \mod{B\Phi F }^2 (0) + \mod{(B\Phi F )'}^2 (0)
 \eec
Using the normalization $F(0)=1$, this  quadratic inequality will give a
superior and an inferior bound on $\rho^2$. 
 
\section{Heavy quark spin symmetry}
 The next step is to try to improve these bounds. Obviously, one way
of achieving this is to take into account as many terms as possible in the
right hand side of (\ref{impi1}).
 It is easily seen that, with some phenomenological input, 
the contribution due to uniparticle states can be computed; thus we get
\cite{raf1}
\bec{piup}
 \Pi'(q^2) \geq  \frac{27}{4\pi \al^2}\ \sum_{i}
  \frac{ M_{\Up_i} \Ga_{\Up_i} }{ (q^2- M^2_{\Up_i}  )^2}
+{1\over 16\pi^2}
 \int_{t_0}^{\infty} \frac{(t-t_0)^{3/2}}{t^{3/2}(t-q^2)^2}\ \mod{F(t)}^2
dt 
\eec
 where the widths $ \Ga_{\Up_i} $ are physically measurable quantities
defined by \cite{raf1}:
  $$ \sigma( e^+ e^- \rightarrow \Up_i )= 12 \pi ^2 \de{t-M^2_ {\Up_i} }
  \frac{ \Ga_{\Up_i} }{M_{\Up_i} }\  $$

Further, we try to include contributions from the $B-\bar{B}^*$,
$B^*-\bar{B}$  and
$B^* -\bar{B}^*$ states. The relevant matrix elements in (\ref{impi1}) can
be expressed through the following form-factors :
 \bec{ffv}
 < B^*(p',\epsilon )\vert V^\mu \vert
B(p)>={2i\epsilon^{\mu\nu\alpha\beta}
\over m_B+m_B^*}\epsilon_\nu p'_\alpha p_\beta V(q^2)
\eec

$$< B^*(p',\epsilon ' )\vert V^\mu \vert B^*(p,
\epsilon)>=F_1(q^2)(\epsilon\cdot \epsilon ')P_\mu+ 
 F_2(q^2)[\epsilon_\mu (\epsilon'\cdot P)
 +\epsilon '_\mu (\epsilon\cdot P)] $$
\bec{fff1234}
 +F_3(q^2){(\epsilon\cdot P) (\epsilon'\cdot P) \over m_{B^*}^2}P_\mu
 +F_4(q^2)[\epsilon_\alpha (\epsilon'\cdot P)-\epsilon'_\alpha
 (\epsilon\cdot P)]
{g^{\mu\alpha} q^2-q^\mu q^\alpha \over m_{B^*}^2}
\eec
(where  $ P=p+p' , q=p-p' $); thus,instead of (\ref{inec1}),(\ref{piup})
 we have :
\bec{incomp}
 \Pi'(q^2) \geq  \frac{27}{4\pi \al^2}\ \sum_{i}
  \frac{ M_{\Up_i} \Ga_{\Up_i} }{ (q^2- M^2_{\Up_i}  )^2} +
  \eec
$${1\over 16\pi^2}  \left\{ \int_{t_0}^{\infty}
 \frac{(t-t_0)^{3/2}}{t^{3/2}(t-q^2)^2}\ \mod{F(t)}^2 dt
+ \int_{t_0^*}^{\infty} \frac{4t}{t_0^*}
\frac{(t-t_0^*)^{3/2}}{t^{3/2}(t-q^2)^2}\ \mod{V(t)}^2 dt +\right. $$
$$\left. \int_{t_0^{**}}^{\infty}
\frac{(t-t_0^{**})^{3/2}}{t^{3/2}(t-q^2)^2}
\left[  2 \mod{F_1(t)}^2+2 \frac{t}{t_0^{**}}
\mod{F_2 (t)}^2+ \mod{\hat{F}_3 (t)}^2 +
\left( \frac{4 t}{t_0^{**}} \right)^2 \mod{F_4(t)}^2 \right] dt \right\}
$$ 

where $t_0^*=(m_B+m_{B^*})^2, t_0^{**}=4m_{B^*}^2$ ; $\hat{F_3}$ is a linear 
combination of $F_1$, $F_2$ and $F_3$ (see (\ref{f3edf}) below).
\footnote{ equation (\ref{incomp}) here differs from eq. (13) 
in \cite{capmac} by a factor 2 in 
the term containing $V(t)$, because in  \cite{capmac} the contribution of
$B^* -\bar{B}$ states was not included}
 
Having six unknown functions in it, this inequality is, in this form, of
no use. At this point, the results of HQET can be of help; it can be shown
that the form-factors are related in the neighborhood of zero-recoil point
as follows: 
\bec{limff}
 V(t),F_2(t) \rightarrow F(t) , F_1(t) \rightarrow -F(t) 
\eec
 $$ F_3,F_4(t) \rightarrow 0 \ , \ \mbox{when~~~~} t= q^2 \rightarrow 0$$ 

 If one assumes \cite{raf} that these relations hold on the
entire unitarity cut then the inequality (\ref{incomp}) can be written in
terms
of a single form factor $F$. But this assumption was shown not to hold, 
especially near thresholds \cite{ball}. 
 
 Treating each term in (\ref{incomp}) separately, we can still obtain an
inequality like (\ref{qin}) :  
\bec{qintot}
 1 \geq \mod{B\Phi F }^2 (0) + \mod{(B\Phi F )'}^2 (0) +
 \mod{B\Phi_V V }^2 (0) + \mod{(B\Phi_V V )'}^2 (0) +\ldots 
\eec
 The full expressions for  $\Phi , \Phi_V ...$ are given in \cite{capmac}
 . This
relation contains only the values of the form factors and of their
derivatives at $q^2 =0$ ; therefore the relations (\ref{limff}) hold, and 
(\ref{incomp}) will become :
\bec{finec}
 a ( \rho^2 )^2 - 2b \rho^2 +c \leq 1 
\eec
 $$ a = 64 \sum (B\Phi)^2(0) \mbox{~~~} 
 b= 8 \sum(B\Phi)(0) (B\Phi)'(0) $$
$$ c = \sum \left[  (B\Phi)^2(0) + (B\Phi)'^2(0) \right] -1$$
with solution :
\bec{sol}
 b-\sqrt{b^2-ac} \leq \rho^2 \leq  b+\sqrt{b^2-ac} 
\eec
This is our final result.

 In deriving (\ref{incomp}),
there is a technical point worth discussing.
The contribution of
$B-\bar{B}$ and $B-\bar{B}^*$ in the spectral sum (\ref{impi1}) are
parametrized by single form factors $F$ and $V$; which appear in the
integral (\ref{incomp}) as $\mod{F}^2$ and $\mod{V}^2$; but the matrix
element of $B^*-\bar{B}^*$ is parametrized by four form factors, which
appear in (\ref{incomp}) as a positive definite quadratic form. To be able
to apply the method described in section 2, we have to write this
quadratic form as
a sum of squares;  it actually appears this way in (\ref{incomp}),
$\hat{F}_3$ being defined as :
\bec{f3edf}
\hat{F}_3 (t) = (\frac{2t}{t_0^{**}} -1) F_1(t) +
\frac{2t}{t_0^{**}} F_2(t) + 
\frac{2t}{t_0^{**}} (\frac{2t}{t_0^{**}} -1)F_3(t)
\eec

\section{Results and discussion}

To obtain numerical results for the bounds (\ref{sol})  there are two
parameters whose value has to be chosen (between certain limits). The first
one is $q^2$ in (\ref{incomp}). Theoretically, we can chose any value for
$q^2$ from $-\infty$ to $m_{\Up_1}^2$; the best results are obtained with
$q^2$ as big as possible. In practice, $\Pi'(q^2)$ is evaluated
\cite{svz,kuhn} (up to three loops)  perturbatively; the reliability of this
evaluation increases if $q^2$ is far from the physical region; it seems
that $q^2=50\ {\rm GeV}^2$ is as close as we can get to $m_{\Up_1}^2 =
89.5 \
{\rm GeV}^2$ and still believe in the perturbation series.
 The second  parameter is the mass of the b quark, which appears in the
evaluation
of $\Pi$. As this quantity is not well defined, we choose to vary it from
$4.7$ to $5\ {\rm GeV}$. Higher mass gives stronger limits.

In previous work, taking into account only the contribution of
$\Up$ and
$B-\bar{B}$ states (not using spin symmetry at all) the following
limits were obtained \cite{raf1} :
$-5.0 \leq \rho^2 \leq 4.5$. Taking into account the
contribution of $B-\bar{B}$ and $B-\bar{B}^*$ the bounds become
 $ -0.90 \leq \rho^2
\leq 2.60 $ \cite{capspin}.These results were obtained with $q^2=0$, $m_b
= 5\ {\rm GeV}$ and using one loop approximation to compute $\Pi'(0)$.
 Because the two
and three loop contribution to $\Pi'(0)$ increase its value by
approximatively 40\% \cite{capmac} (!), these limits are actually stronger
than what they
should be. 
   
 Using the full apparatus presented in this 
 paper, we obtain   : at $q^2=0$:
$ -0.2 \leq \rho^2 \leq 1.85$ for $m_b=4.7 \ {\rm GeV}$; 
$ -0.1 \leq \rho^2 \leq 1.76$ for $m_b=5.0 \ {\rm GeV}$;
  at $q^2=50\ {\rm GeV}^2$ :
$ -0.0 \leq \rho^2 \leq 1.6$ ,$m_b=4.7 \ {\rm GeV}$ and
$ 0.3 \leq \rho^2 \leq 1.2$ for $m_b=5.0 \ {\rm GeV}$.

These results show that using only general properties of QCD (as
analyticity and unitarity) we can derive nontrivial constraints on the
behavior of heavy mesons form factors at transfer  momentum close to
zero. The use of heavy quark spin symmetry brings significant
improvements.

Further results are expected from the application of this
method to other transitions, like  the physically interesting case of $B$
to $D^*$ transition \cite{capneu}, \cite{bgl}.

\section{Acknowledgements}
 This paper presents results obtained in collaboration with Irinel Caprini. A more 
 detailed (and technical) presentation can be found 
 in \cite{capmac}. I would also like to thank I. 
Caprini for helpful suggestions in preparing this paper,
as well as for introducing me to this subject.


\section{References}
\begin{thebibliography}{20}
\bibitem{isgur} N.Isgur and M.B.Wise, Phys.Lett.{\bf B232}(1989) 113;
{\bf B237}(1990) 523.
\bibitem{review} M. Neubert, Phys. Rep. {\bf 245}, 259 (1994).
\bibitem{ball} P.Ball, H.G.Dosch, M.A.Shifman, Phys.Rev.D 47, 4077 (1993),
hep-ph/9211296.
\bibitem{bart} G.Barton, { \it Introduction to dispersion techniques in
quantum field theory}, New York W.A.Benjamin 1965.
\bibitem{duren} P.Duren, { \it Theory of $H^p$ Spaces},
New York: Academic Press, 1970.
\bibitem {raf} E.de Rafael and J.Taron, Phys.Lett. B{\bf 282}, 215 (1992).
\bibitem {raf1} E.de Rafael and J.Taron, Phys.Rev.D{\bf 50}, 373  (1994), 
hep-ph/9306214.
\bibitem{svz} M.A.Shifman, A.I. Vainshtein, and V.I. Zakharov, Nucl. Phys. 
{\bf B147}, 385 (1979); {\bf B147}, 448 (1979) 
\bibitem{kuhn} K.G.Chetyrkin, J.H.K\"uhn, M.Steihauser, Phys.Lett.
{\bf B371},93 (1996).
\bibitem{capspin} I.Caprini Phys.Rev.D{\bf 53}, 4082 ( 1996).
\bibitem{capmac} I.Caprini and C.Macesanu, Phys.Rev.D{\bf 54}, 5686
(1996), hep-ph/9605365.
\bibitem{capneu}I. Caprini, L. Lellouch
and M. Neubert, preprint CERN-TH/97-91,
CPT-97/P3480, May 1997.
\bibitem{bgl} C.G.Boyd, B.Grinstein and R.F.Lebed, preprint CMU-HEP 97-07,
 UCSD/PTH 97-12, May 1997,  hep-ph/9705252
\end{thebibliography}
  \end{document}